\documentclass[twocolumn,prl]{revtex4}
\usepackage{float}
\usepackage{makeidx}
\usepackage{graphicx}

\begin{document}

\title{Continuous-Variable Quantum Cloning of Coherent states with Phase-Conjugate Input Modes Using
Linear Optics}
\author{Haixia Chen, Jing Zhang$^{\dagger }$}
\affiliation{State Key Laboratory of Quantum Optics and Quantum
Optics Devices, Institute of Opto-Electronics, Shanxi University,
Taiyuan 030006, P.R.China \label{in}}

\begin{abstract}
We propose a scheme for continuous-variable quantum cloning of
coherent states with phase-conjugate input modes using linear
optics. The quantum cloning machine yields $M$ identical optimal
clones from $N$ replicas of a coherent state and $N$ its replicas of
phase conjugate. This scheme can be straightforwardly implemented
with the setup accessible at present since its optical
implementation only employs simple linear optical elements and
homodyne detection. Compared with the original scheme for continuous
variables quantum cloning with phase-conjugate input modes proposed
by Cerf and Iblisdir [Phys. Rev. Lett. 87, 247903 (2001)], which
utilized a nondegenerate optical parametric amplifier, our scheme
loses the output of phase-conjugate clones and is regarded as
irreversible quantum cloning.
\end{abstract}

\maketitle

\section{Introduction}

Quantum cloning plays an important role in quantum information and
quantum communication. It has been shown that quantum cloning might
improve the performance of some computational tasks \cite{one} and
it is believed to be the optimal eavesdropping attack for a certain
class of quantum cryptography \cite{two}. It also opens an avenue
for understanding the concepts of quantum mechanics and measurement
theory further. So the quantum cloning which achieves the optimal
cloning transformation compatible with the quantum no cloning
theorem has always being a hot research topic. Such a quantum
cloning machine was first considered by Buzek and Hillery for qubits
\cite{three} and later extended to the continuous-variable (CV)
regime by Cerf et al. \cite{four}. CV quantum cloning has been
extensively studied in the recent years for the relative ease in
preparing and manipulating quantum states. The theoretical proposals
for the experimental implementations of CV quantum cloning have been
proposed \cite{five,six,seven,seven1}.

A recent result in the context of measurement has revealed that more
quantum information can be encoded in the antiparallel pairs of
spins than parallel pairs \cite{eight}. Subsequently, a result that
a pair of conjugate Gaussian states can carry more information than
by using the same states twice has been extended to continuous
variables \cite{nine}. This result makes it possible to yield better
fidelity with the cloning machine admitting antiparallel input
qubits or phase-conjugate input modes thereby opening a new avenue
in the investigation of quantum cloning. Based on the above
properties, Cerf and Iblisdir put forward a CV cloning
transformation \cite {ten} that takes as input $N$ replicas of a
coherent states and $N^{^{\prime }}$ replicas of its complex
conjugate, and produces $M$ optimal clones of the coherent state and
$M^{\prime }=M+N^{\prime }-N$ phase-conjugate clones (anticlones, or
time reversed states). This is the first scheme for the
phase-conjugate input (PCI) cloner of continuous variables. It is,
nonetheless, difficult for the practical experimental realization of
the proposed PCI cloner due to the difficulties associated with the
physical implementation of the optical parametric amplifier.
Recently, a much simpler but efficient CV quantum cloning machine
based on linear optics and homodyne detection was proposed and
realized experimentally by Andersen et al. \cite {eleven}. Later,
this protocol is extended to various quantum cloning cases, such as
asymmetric cloning \cite{eleven1} and so on \cite{twelve}. According
to classifying the quantum clone to irreversible and reversible
types in the perspective of quantum information distribution
\cite{teleclone}, the quantum cloning with linear optics
\cite{eleven} is local and irreversible, in which the anticlones are
lost. Perfect distribution do not allow losing any the quantum
information of the transmitted unknown state, that means this
process is reversible and the unknown state can be reconstructed in
a quantum system again.

In this paper, we propose a protocol of CV quantum cloning of
coherent states with phase-conjugate input modes using linear
optics. The $N+N\rightarrow M$ quantum cloning machine yields $M$
identical optimal clones from $N$ replicas of a coherent state and
$N$ its replicas of phase conjugate. This scheme is regarded as
local and irreversible PCI quantum cloning because the anticlones
are lost. We also show that $N+N\rightarrow M$ irreversible PCI
quantum cloning machine may be changed into $N+N\rightarrow M+M$
reversible PCI quantum cloning machine by the introduction of an EPR
(Einstein-Podolsky-Rosen) entangled ancilla. It shows that the
optimal fidelity of the anticlones requires the maximally EPR
entangled state.

\section{$1+1\rightarrow M$ Irreversible PCI Quantum Cloning}

The quantum states we consider in this paper are described with the
electromagnetic field annihilation operator $\hat{a}=(\hat{X}+i\hat{P})/2$,
which is expressed in terms of the amplitude $\hat{X}$ and phase $\hat{P}$
quadrature with the canonical commutation relation $[\hat{X},\hat{P}]=2i$.
Without any loss of generality, the quadrature operators can be expressed in
terms of a steady state and fluctuating component as $\hat{A}=\langle \hat{A}%
\rangle +\Delta \hat{A}$, which have variances of $V_A=\langle \Delta \hat{A}%
^2\rangle $ ($\hat{A}=\hat{X}$ or $\hat{P})$. The input coherent
state and its phase-conjugate state to be cloned will be described
by $\left| \alpha _{in}\right\rangle =\left| \frac 12\left(
x_{in}+ip_{in}\right) \right\rangle $ and $\left| \alpha
_{in}^{*}\right\rangle =\left| \frac 12\left( x_{in}-ip_{in}\right)
\right\rangle $ respectively, where $x_{in}$
and $p_{in}$ are the expectation values of $\hat{X}_{in}$ and $\hat{P}_{in}$%
. The cloning machine generates many clones of input state characterized by
the density operator $\hat{\rho}_{clone}$ and the expectation $x_{clone}$
and $p_{clone}$. The quality of the cloning machine can be quantified by the
fidelity, which a overlap between the input state and the output state. It
is defined by \cite{fifteen}

\begin{eqnarray}
F &=&\left\langle \alpha _{in}\right| \hat{\rho}_{clone}\left|
\alpha _{in}\right\rangle =\frac 2{\sqrt{( 1+\Delta
^2\hat{X}_{clone})
( 1+\Delta ^2\hat{P}_{clone}) }}  \nonumber \\
&&*\exp \left[ -\frac{( x_{clone}-x_{in}) ^2}{2(1+\Delta ^2%
\hat{X}_{clone}) }-\frac{( y_{clone}-y_{in}) ^2}{2( 1+\Delta
^2\hat{P}_{clone}) }\right]  \label{1}
\end{eqnarray}
In the case of unity gains, i.e., $x_{clone}=x_{in}$, the fidelity is
strongly peaked and changed into

\begin{equation}
F=\frac 2{\sqrt{( 1+\Delta ^2\hat{X}_{clone}) ( 1+\Delta ^2%
\hat{P}_{clone}) }}.  \label{2}
\end{equation}

%
\begin{figure}
\centerline{
\includegraphics[width=3.3in]{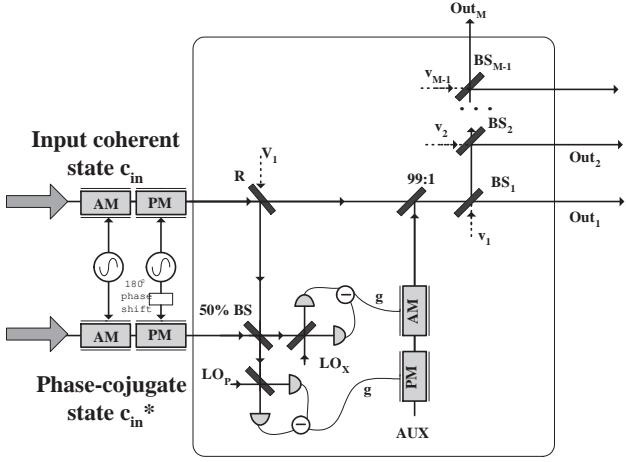}
} \vspace{0.1in}
\caption{ A schematic diagram of $1+1\rightarrow M$ irreversible PCI
quantum cloning. BS: Beam splitter, LO: Local oscillator, AM:
Amplitude modulator, PM: Phase modulator and AUX: Auxiliary beam.
\label{Fig1} }
\end{figure}

Let us first illustrate the protocol in the simplest case of
$N=N^{\prime }=1 $ as shown in Fig. 1. The input coherent state
$\hat{c}_{in}$ and its phase-conjugate state $\hat{c}_{in}^{*}$ are
prepared by an amplitude modulator and a phase modulator
respectively. The modulated signals on the amplitude modulators are
in-phase and the modulated signals on the phase modulators are
out--of-phase. The input mode $\hat{c}_{in}$ is divided by a
variable beam splitter with transmission rate $T$ and reflectivity
rate $R$. The reflected output
$\hat{c}_{1r}=\sqrt{R}\hat{c}_{in}+\sqrt{T}\hat{V}_1$, where the
annihilation operator $\hat{V}_1$ represents the vacuum mode
entering the beam splitter, is combined with its phase-conjugate state $\hat{%
c}_{in}^{*}$ at a 50/50 beam splitter. Then we perform homodyne measurements
on the two output beams to achieve the amplitude and the phase quadratures
simultaneously. The measured quadratures are

\begin{eqnarray}
\hat{X}_m &=&\frac 1{\sqrt{2}}( \sqrt{R}\hat{X}_{c_{in}}+\sqrt{T}\hat{X}%
_{V_1}+\hat{X}_{c_{in}^{*}})  \nonumber \\
\hat{P}_m &=&\frac 1{\sqrt{2}}( \sqrt{R}\hat{P}_{c_{in}}+\sqrt{T}\hat{P}%
_{V_1}-\hat{P}_{c_{in}^{*}}) .  \label{3}
\end{eqnarray}
We use the measurement outcomes to modulate the amplitude and phase
of an auxiliary coherent beam via two independent modulators with a
scaling factor $g $ \cite{sixteen}. This beam is then combined at a
99/1 beam splitter with the transmitted part of mode $\hat{c}_{in}$,
hereby displacing this part according to the measurement outcomes
\cite{sixteen}.
Corresponding to the transformation $\hat{A}\rightarrow \hat{D}^{+}\hat{A}%
\hat{D}=\hat{A}+\left( \hat{X}_m+i\hat{P}_m\right) /2$ in the Heisenberg
representation, the displaced field can be expressed as

\begin{eqnarray}
\hat{c}_{disp} &=&( \sqrt{1-R}+\frac g{\sqrt{2}}\sqrt{R})
\hat{c}_{in}  \nonumber \\
&&-( \sqrt{R}-\frac g {\sqrt{2}}\sqrt{1-R}) \hat{V}_1+\frac g
{\sqrt{2}}\hat{c}_{in}^{*+}  \label{5}
\end{eqnarray}
where $\hat{c}_{disp}$ is the annihilation operator for the
displaced field. By choosing $g =\sqrt{2R/\left( 1-R\right) }$, we
can cancel the vacuum noise of the displaced field. Then the
displaced field is given by

\begin{equation}
\hat{c}_{disp}^c=\frac 1{\sqrt{1-R}}\hat{c}_{in}+\frac{\sqrt{R}}{\sqrt{1-R}}%
\hat{c}_{in}^{*+}.  \label{disp1}
\end{equation}
We can see Eq. (\ref{disp1}) equal to a phase-insensitive
amplification with gain $G=\frac 1{1-R}$.

In the final step the displaced field is distributed into $M$ clones $\{\hat{%
a}_l^{^{\prime }}\}$ $(l=1,2,\cdots M$ $)$ by a sequence of $M-1$ beam
splitters with appropriately adjusted transmittances and reflectances. Then
the output of cloning machine can be expressed by

\begin{eqnarray}
\hat{a}_1^{^{\prime }} &=&\sqrt{\frac 1M}\hat{c}_{disp}^c+\sqrt{\frac{M-1}M}%
\hat{v}_1  \nonumber \\
\hat{a}_2^{^{\prime }} &=&\sqrt{\frac 1M}\hat{c}_{disp}^c-\sqrt{\frac
1{M\left( M-1\right) }}\hat{v}_1+\sqrt{\frac{M-2}{M-1}}\hat{v}_2  \nonumber
\\
&&\cdots  \nonumber \\
\hat{a}_{M-1}^{^{\prime }} &=&\sqrt{\frac 1M}\hat{c}_{disp}^c-\sqrt{\frac
1{M\left( M-1\right) }}\hat{v}_1-\sqrt{\frac 1{\left( M-1\right) \left(
M-2\right) }}  \nonumber \\
&&*\hat{v}_2-\cdots -\sqrt{\frac 1{3*2}}\hat{v}_{\left( M-2\right) }+\sqrt{%
\frac 12}\hat{v}_{\left( M-1\right) }  \nonumber \\
\hat{a}_M^{^{\prime }} &=&\sqrt{\frac 1M}\hat{c}_{disp}^c-\sqrt{\frac
1{M\left( M-1\right) }}\hat{v}_1-\sqrt{\frac 1{\left( M-1\right) \left(
M-2\right) }}  \nonumber \\
&&*\hat{v}_2-\cdots -\sqrt{\frac 1{3*2}}\hat{v}_{\left( M-2\right) }-\sqrt{%
\frac 12}\hat{v}_{\left( M-1\right) }  \label{7}
\end{eqnarray}
where $\hat{v}_k\left( k=1,2,\cdots M-1\right) $ refer to the
annihilation operators of vacuum mode entering the $BS_1,BS_2,\cdots
BS_{M-1}$ respectively. Eq. (\ref{7}) shows that each output mode
contains the in the
displaced field $\hat{c}_{disp}^c$ with a factor of $1/\sqrt{M}$. Note that both terms $%
\hat{c}_{in}$ and $\hat{c}_{in}^{*+}$ in the Eq. (\ref{disp1})
contribute to the total coherent signal with a factor of
$1/\sqrt{1-R}+\sqrt{R}/\sqrt{1-R}$ and noise variances with
$\sqrt{1+R}/\sqrt{1-R}$ in the output $\hat{c}_{disp}^c$. Since each
output cloner should includes one unit of the input coherent state,
the $R$ must satisfy

\begin{equation}
\frac 1{\sqrt{1-R}}+\frac{\sqrt{R}}{\sqrt{1-R}}=\sqrt{M}.  \label{8}
\end{equation}
The $R$ can be easily determined by solving the above equation and given by

\begin{equation}
R=\frac{\left( M-1\right) ^2}{\left( M+1\right) ^2}.  \label{9}
\end{equation}
According to the Eqs. (\ref{disp1},\ref{7},\ref{9}), the variances
of the clones can be written as

\begin{eqnarray}
\Delta ^2\hat{X}_{a_l^{^{\prime }}} &=&\Delta
^2\hat{P}_{a_l^{^{\prime
}}}=\frac 1M\frac{1+R}{1-R}+\frac{M-1}M  \label{10} \\
&=&1+\frac{\left( M-1\right) ^2}{2M^2}.  \nonumber
\end{eqnarray}
The fidelity can be get through Eq. (\ref{2})

\begin{equation}
F_{\left( _1^1\right) \rightarrow M}=\frac{4M^2}{4M^2+\left( M-1\right) ^2}.
\label{11}
\end{equation}
This procedure is optimal clearly to produce $M$ clones. Now we
compare the fidelity of $M$ clones from the phase-conjugate input
modes with from the two identical replicas. The fidelity of the
standard 2-to-M cloning are given by \cite{seventeen}

\begin{equation}
F_{2\rightarrow M}=\frac{2M}{3M-2}  \label{12}
\end{equation}
In the special case $M=2$, the standard cloning can be achieved
perfectly with fidelity equal to one while the phase-conjugate
cloner yields an additional variance which will lead to a lower
fidelity. It is, nonetheless, obvious that phase-conjugate cloner
yields better fidelity than the standard cloning when $M\geq 3$. In
the limit of large $M\rightarrow \infty $, we could see $F_{\left(
_1^1\right) \rightarrow \infty }=\frac 45$ compared with the
standard cloning $F_{2\rightarrow \infty }=\frac 23$. This shows
that more information can be encoded into a pair of conjugate
Gaussian states than by using the two same states, which has been
shown in Ref. \cite {nine}. Compared with the original scheme for
continuous variables quantum cloning with phase-conjugate input
modes proposed by Cerf and Iblisdir \cite {ten}, which utilized a
nondegenerate optical parametric amplifier, our scheme loses the
anticlones and is regarded as irreversible PCI\ quantum cloning.

Now we consider the realistic conditions where the homodyne detector
efficiency is not unity. If $\eta$ expresses the homodyne detector
efficiency, the measured amplitude and the phase quadratures are
give by

\begin{eqnarray}
\hat{X}_m &=&\sqrt{\frac {\eta}{2}}( \sqrt{R}\hat{X}_{c_{in}}+\sqrt{T}\hat{X}%
_{V_1}+\hat{X}_{c_{in}^{*}})+\sqrt{1-\eta}\hat{X}_{V_{D1}}  \nonumber \\
\hat{P}_m &=&\sqrt{\frac {\eta}{2}}( \sqrt{R}\hat{P}_{c_{in}}+\sqrt{T}\hat{P}%
_{V_1}-\hat{P}_{c_{in}^{*}})+\sqrt{1-\eta}\hat{P}_{V_{D2}}
\label{meas}
\end{eqnarray}
where $\hat{X}_{V_{D1}}$ and $\hat{P}_{V_{D2}}$ are the vacuum noise
introduced from the losses of the homodyne detector. With the
measured results, the displaced field can be expressed as
\begin{eqnarray}
\hat{c}_{disp} &=&( \sqrt{1-R}+g\sqrt{\frac {\eta}{2}}\sqrt{R})
\hat{c}_{in}-( \sqrt{R}-g\sqrt{\frac {\eta}{2}}\sqrt{1-R})
\hat{V}_1  \nonumber \\
&&+g\sqrt{\frac
{\eta}{2}}\hat{c}_{in}^{*+}+\sqrt{1-\eta}g\hat{X}_{V_{D1}}+\sqrt{1-\eta}g\hat{P}_{V_{D2}}
\label{dispp}
\end{eqnarray}
By choosing $g =\sqrt{2R/\eta( 1-R) }$, the displaced field is given
by
\begin{eqnarray}
\hat{c}_{disp}^c&=&\frac 1{\sqrt{1-R}}\hat{c}_{in}+\frac{\sqrt{R}}{\sqrt{1-R}}%
\hat{c}_{in}^{*+}\nonumber \\
&&+\sqrt{\frac{2R(1-\eta)}{(1-R)\eta}}(\hat{X}_{V_{D1}}+\hat{P}_{V_{D2}}).
\label{dispp1}
\end{eqnarray}
According to the Eqs. (\ref{7},\ref{8},\ref{9}), the variances of
the clones can be written as

\begin{eqnarray}
\Delta ^2\hat{X}_{a_l^{^{\prime }}} =\Delta ^2\hat{P}_{a_l^{^{\prime
}}}&=&\frac 1M\frac{1+R}{1-R}+\frac 1M\frac{2R(1-\eta)}{(1-R)\eta}  \label{qua} \\
+\frac{M-1}M &=&1+\frac {1}{\eta}\frac{\left( M-1\right) ^2}{2M^2}.
\nonumber
\end{eqnarray}
The fidelity can be get through Eq. (\ref{2})
\begin{equation}
F_{\left( _1^1\right) \rightarrow M}=\frac{4\eta M^2}{4\eta
M^2+\left( M-1\right) ^2}. \label{f}
\end{equation}
It clearly shows that the fidelity of the clones is degraded due to
the losses of the homodyne detection.

\section{$N+N\rightarrow M$ Irreversible PCI Quantum Cloning}

We now generalize $1+1\rightarrow M$ case to $N+N\rightarrow M$ irreversible
PCI quantum cloning, which produces $M$ clones from input $N$ replicas of a
coherent states and $N$ replicas of its complex conjugate as illustrated in
Fig. 2. First, we concentrate $N$ identically prepared coherent states $%
\left| \Phi \right\rangle $ described by $\left\{ \hat{a}_l\right\} $ $%
(l=1,\cdots N)$ into a single spatial mode $\hat{c}_1$ with amplitude $\sqrt{%
N}\Phi $. This operation can be performed by interfering $N$ input modes in $%
N-1$ beam splitters, which yields the mode

%
\begin{figure}
\centerline{
\includegraphics[width=3.3in]{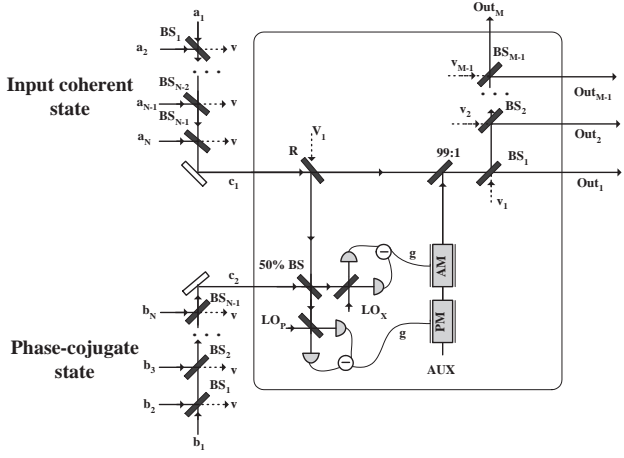}
} \vspace{0.1in}
\caption{ A schematic diagram of $N+N\rightarrow M$ irreversible PCI
quantum cloning. \label{Fig2} }
\end{figure}

\begin{equation}
\hat{c}_1=\frac 1{\sqrt{N}}\sum_{l=1}^N\hat{a}_l  \label{13}
\end{equation}
and $N-1$ vacuum modes. The same method can be used on the generation of the
phase-conjugate input mode $\hat{c}_2$ with amplitude $\sqrt{N}\Phi ^{*}$
from the $N\ $replicas of $\left| \Phi ^{*}\right\rangle $ stored in the $N\
$modes $\left\{ b_l\right\} $ $(l=1,\cdots N)$, which is expressed as
\begin{equation}
\hat{c}_2=\frac 1{\sqrt{N}}\sum_{l=1}^N\hat{b}_l.  \label{14}
\end{equation}
Then, $\hat{c}_1$ and $\hat{c}_2$ is transported into the cloning
machine same as Fig. 1. The displaced field is given by

\begin{equation}
\hat{c}_{disp}^c=\frac 1{\sqrt{1-R}}\hat{c}_1+\frac{\sqrt{R}}{\sqrt{1-R}}%
\hat{c}_2^{+}.  \label{disp2}
\end{equation}
The terms $\hat{c}_1$ and $\hat{c}_2^{+}$ in the Eq. (\ref{disp2})
contribute
to the total coherent signal with a factor of $\sqrt{N}(1/\sqrt{1-R}+\sqrt{R}%
/\sqrt{1-R})$ and noise variances with $(1+R)/(1-R)$ in the output $\hat{c}%
_{disp}^c$. Since each output cloner should includes one unit of the input
coherent state, the $R$ must satisfy

\begin{equation}
\sqrt{N}(\frac 1{\sqrt{1-R}}+\frac{\sqrt{R}}{\sqrt{1-R}})=\sqrt{M}.
\end{equation}
The $R$ can be easily determined by solving the above equation and given by

\begin{equation}
R=\frac{\left( M-N\right) ^2}{\left( M+N\right) ^2}.
\end{equation}
The variance and fidelity of $\left( _N^N\right) \rightarrow $ $M$\
cloner will be given by

\begin{equation}
\Delta ^2\hat{X}_{a_l^{^{\prime }}}=\Delta ^2\hat{P}_{a_l^{^{\prime }}}=1+%
\frac{\left( M-N\right) ^2}{2M^2N},  \label{16}
\end{equation}

\begin{equation}
F_{\left( _N^N\right) \rightarrow M}=\frac{4M^2N}{4M^2N+\left( M-N\right) ^2}%
.  \label{18}
\end{equation}
Obviously, Eqs. (\ref{10}) and (\ref{11}) can be obtained by Eqs.
(\ref{16}) and (\ref{18}) for $N=N^{\prime }=1$. The result also
coincides with that obtained in Ref. \cite{ten}. However, the output
anticlones are lost in this scheme. The advantage of dealing with
$N\ $pair of complex conjugate inputs can still be most easily
illustrated in the limit of infinite number of clones, $M\rightarrow
\infty $ , from Eq. (\ref{18}) we get $F_{_{\left( _N^N\right)
\rightarrow M}}=\frac{4N}{4N+1}$ while the standard cloning machine
fidelity $F_{2N\rightarrow M}=\frac{2N}{2N+1}$.

\section{Reversible PCI Cloning with Linear Optics and EPR Entanglement}

A scheme for a phase conjugating amplifier with the nonlinearity put
off-line was proposed \cite{eighteen}. Employing this protocol, we show that $%
N+N\rightarrow M$ irreversible PCI quantum cloning machine as shown
in Fig. 2 become $N+N\rightarrow M+M$ reversible PCI quantum cloning
machine by the introduction of an EPR entangled ancilla (two-mode
Gaussian entangled state) as shown Fig. 3. One half of the entangled
ancilla is injected into the empty port of the variable beam
splitter. Since the noises injected into the empty port of the
variable beam splitter are canceled in the displaced field, the
displaced field don't depend on the injected noises. Thus the above
results for the clones are always valid. The other half of the
entangled ancilla is also displaced according to the classical
measurement outcomes with a scaling factor $g_{1}$ and expressed as

%
\begin{figure}
\centerline{
\includegraphics[width=3.3in]{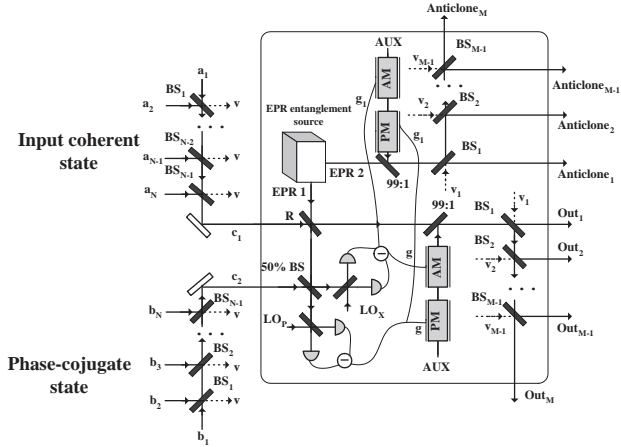}
} \vspace{0.1in}
\caption{ A schematic diagram of reversible PCI cloning with linear
optics and EPR entanglement. \label{Fig3} }
\end{figure}

\begin{equation}
\hat{e}_{disp}=\frac {g_{1}}{\sqrt{2}}\sqrt{R}\hat{c}_1^{+}+\frac {g_{1}}{\sqrt{2}}\sqrt{%
1-R}\hat{b}_{EPR1}^{+}+\hat{b}_{EPR2}+\frac
{g_{1}}{\sqrt{2}}\hat{c}_2. \label{EPRdisp}
\end{equation}
By choosing $g_{1}=\sqrt{2/\left( 1-R\right) }$, the displaced EPR
beam is given by

\begin{equation}
\hat{e}_{disp}=\frac{\sqrt{R}}{\sqrt{1-R}}\hat{c}_1^{+}+\frac 1{\sqrt{1-R}}%
\hat{c}_2+(\hat{b}_{EPR1}^{+}+\hat{b}_{EPR2}).
\end{equation}
The EPR entangled beams $\hat{b}_{EPR1}$, $\hat{b}_{EPR2}$ have the
very strong correlation property, such as both their sum-amplitude
quadrature variance $\langle \Delta
(\hat{X}_{b_{EPR1}}+\hat{X}_{b_{EPR2}})^2\rangle
=2e^{-2r}$, and their difference-phase quadrature variance $\langle \Delta (%
\hat{Y}_{b_{EPR1}}-\hat{Y}_{b_{EPR2}})^2\rangle =2e^{-2r}$, are less
than quantum noise limit. In the final step the displaced EPR beam
is distributed into $M$ anticlones $\{\hat{b}_l^{^{\prime }}\}$
$(l=1,2,\cdots M$ $)$ by a sequence of $M-1$ beam splitters with
appropriately adjusted transmittances and reflectances. The
expression of the output anticlones is similar to Eq. \ref{7}. The
variance and fidelity of anticloner will be given by

\begin{equation}
\Delta ^2\hat{X}_{b_l^{^{\prime }}}=\Delta ^2\hat{P}_{b_l^{^{\prime }}}=1+%
\frac{\left( M-N\right) ^2}{2M^2N}+\frac{2e^{-2r}}M,
\end{equation}

\begin{equation}
F_{\left( _N^N\right) \rightarrow M}^{anti}=\frac{4M^2N}{4M^2N+\left(
M-N\right) ^2+4MNe^{-2r}}.
\end{equation}
It clearly shows that the optimal fidelity of the anticlones
requires the maximally EPR entangled state $r\rightarrow \infty $.
Clearly the reversible PCI cloning with linear optics and EPR
entanglement is equivalent to the original scheme for CV PCI quantum
cloning proposed by Cerf and Iblisdir \cite {ten}, which utilized a
nondegenerate optical parametric amplifier.

\section{Conclusion$\ $}

In conclusion, we have proposed a much simpler and experimentally
feasible continuous variables cloning machine of coherent states
with phase-conjugate inputs using linear optics. Compared with the
original scheme for continuous variables quantum cloning with
phase-conjugate input modes proposed by Cerf and Iblisdir, which
utilized a nondegenerate optical parametric amplifier, our scheme
loses the output of phase-conjugate clones and is regarded as
irreversible quantum cloning. The protocols described here can be
used in various quantum communication protocols, e.g., for the
optimal eavesdropping of a quantum key distribution scheme.

$^{\dagger }$Corresponding author's email address: jzhang74@yahoo.com,
jzhang74@sxu.edu.cn

\section{\textbf{ACKNOWLEDGMENTS}}

\textbf{\ }J. Zhang thanks K. Peng, C. Xie, T. Zhang, and J. Gao for
the helpful discussions. This research was supported in part by
National Fundamental Research Natural Science Foundation of China
(Grant No. 2006CB0L0101), National Natural Science Foundation of
China (Grant No. 60678029), Program for New Century Excellent
Talents in University (Grant No. NCET-04-0256), Doctoral Program
Foundation of Ministry of Education China (Grant No. 20050108007),
the Cultivation Fund of the Key Scientific and Technical Innovation
Project, Ministry of Education of China (Grant No. 705010), Program
for Changjiang Scholars and Innovative Research Team in University,
Natural Science Foundation of Shanxi Province (Grant No.
2006011003), and the Research Fund for the Returned Abroad Scholars
of Shanxi Province.

\end{document}